\documentstyle [rawfonts,somedefs,epsf,psfig]{europhys}
\def\etal{{\hbox{{\tenit\ et al.\/}\tenrm :\ }}}

\def\stars{\bigskip\centerline{***}\medskip}

\newif\ifboo \boofalse

\def\Review#1{\boofalse{\it #1},}

\def\Vol#1{\ifboo Vol. {\bf #1}\else{\bf #1}\fi}
\def\Year#1{\ifboo #1\else(#1)\fi}

\def\Page#1{\ifboo {\rm p. #1}\else{\rm #1}\fi}

\newcommand{\ie}{{\it i.e. \/}}
\newcommand{\eg}{{\it e.g. \/}}

\newcommand{\fig}[1]{Fig.~\ref{#1}}
\newcommand{\eq}[1]{Eq.~(\ref{#1})}

\renewcommand{\vec}[1]{{\bf #1}}
\newcommand{\tens}[1]{{\mathbf #1}}

\newcommand{\gt}{>}

\newcommand{\gtrsim}{\stackrel{>}{\sim}}

\newcommand{\lesssim}{\stackrel{<}{\sim}}

\newlength{\fboxwidth}                 
\begin {document}
\euro{}{}{}{}
\Date{}
\shorttitle{R. RZEHAK \etal DYNAMICS OF STRONGLY DEFORMED POLYMERS IN SOLUTION}
\title{Dynamics of Strongly Deformed Polymers in Solution}
\author{Roland Rzehak  and Walter Zimmermann }
\institute{
Theoretische Physik, 
Universit\"at des Saarlandes,
D-66041 Saarbr\"ucken \\
Institut f\"ur  Festk\"orperforschung and FORUM,
Forschungszentrum J\"ulich,
D-52425 J\"ulich 
}
\rec{}{}
\pacs{
  \Pacs{83}{10.Mj}{Molecular dynamics, Brownian dynamics }
  \Pacs{36}{20.Ey}{Conformation (Statistics and Dynamics)}  
  \Pacs{47}{+d}{Non-Newtonian fluid flows}
}
\maketitle
%
%
\begin{abstract}
Bead spring models for polymers in solution are nonlinear if
either the finite extensibility of the polymer,  
excluded volume effects or 
hydrodynamic interactions between polymer segments 
are taken into account. 
For such models we use a powerful method 
for the determination of the complete relaxation spectrum
of fluctuations at {\it steady state}.
In general, the spectrum and modes differ significantly 
from those of the linear Rouse model.
For a tethered polymer in uniform flow
the  differences are mainly caused by
an inhomogeneous distribution of  tension along the chain
and are most pronounced due to the finite chain extensibility. 
Beyond the dynamics of steady state fluctuations
we also investigate the nonlinear response of the polymer 
to a {\em large sudden change} in the flow.
This response exhibits several distinct regimes with 
characteristic decay laws and shows features which are 
beyond the scope of single mode theories such as the dumbbell model. 
\end{abstract}
%
%
\section{Introduction}
The slow internal dynamics of polymers
makes polymer solutions viscoelastic
which is manifest in such spectacular effects as
turbulent drag reduction. 
In spite of active research for more than 60 years already, 
the understanding of these phenomena is still incomplete 
\cite{Ferry:VEPP-80+Bird:DPL12-87,Gennes:SCPP-81,Doi:TPD-86,Gyr:DRTF-95,Rouse:JCP21-53-1272,Zimm:JCP24-56-269,Gennes:JCP60-74-5030,Hinch:PF20-77-S22}.
Progress towards 
a microscopically founded rather than empirical 
description of flowing polymer solutions 
is possible only by 
a thorough analysis of 
the interplay between 
single polymers and the flow. 
However, only recent experiments on individual DNA molecules allow to 
observe the molecular conformations 
\cite{Perkins:SCI264-94-822,Manneville:EL36-96-413,Quake:NAT388-97-151,Perkins:SCI276-97-2016+Smith:SCI283-99-1724}
and computer simulations facilitate a treatment of 
the nonlinear interactions within the polymer-solvent system 
\cite{Larson:PRE55-97-1794,Hatfield:PRL82-99-3548,Rzehak:NN2-99,Rzehak:NN1-99,Rzehak:NN0-99}. 

The dynamics of fluctuations of a polymer in flow 
is commonly characterized by relaxation times and modes, 
which also determine the linear viscoelastic properties.
These quantities are known analytically 
only for the linear Rouse and Zimm models 
\cite{Rouse:JCP21-53-1272,Zimm:JCP24-56-269}.
Questions arising are:
Is there an effective method for determining 
the complete relaxation spectrum and the corresponding modes 
for fluctuations around an arbitrary stationary state 
also for more realistic {\em nonlinear} polymer models? 
What is their dependence on the polymer deformation and
which decay laws describe the whole relaxation process 
of a strongly stretched polymer to equilibrium? 

Here we give answers for 
the model problem of a tethered polymer in uniform flow
which has been much investigated recently 
both experimentally 
\cite{Perkins:SCI264-94-822,Manneville:EL36-96-413}
and theoretically using blob models
\cite{Brochard:EL26-94-511+Brochard:EL30-95-387+Marciano:MM28-95-985,Rzehak:NN2-99}
and bead spring models 
\cite{Larson:PRE55-97-1794,Hatfield:PRL82-99-3548,Rzehak:NN2-99,Rzehak:NN1-99,Rzehak:NN0-99}.
By simulation of bead spring chains we determine
the relative importance for the polymer dynamics
of various interactions such as 
finite extensibility,
excluded volume interactions ({\bf EVI}), 
and hydrodynamic interactions ({\bf HI}).

Due to Brownian motion, even in a steady state the beads
fluctuate around some time-independent average positions.
In the equilibrium state of a tethered polymer 
these average positions all coincide with the tether-point,
while if the polymer is stretched by a flow or force
they assume non-trivial values. 
Our analysis of the conformational fluctuations in such a statistically steady state 
is based on the covariance matrix of bead positions.
This approach yields 
the full relaxation spectrum
rather than only the longest relaxation time considered in previous works
and it provides numeric values for the relaxation times
which cannot be obtained from common scaling arguments.
Furthermore it gives a corresponding complete set of 
{\it uncorrelated} relaxation modes 
and thus constitutes
a generalization of the normal mode analysis 
used for the linear models of Rouse and Zimm
which is applicable to {\it nonlinear} polymer models.

Theories of polymer deformation in simple shear or elongational flows 
rely on the so-called time criterion \cite{Gennes:JCP60-74-5030}:
the polymer will be significantly deformed when 
the inverse shear or elongation rate $\kappa^{-1}$
is shorter than the longest polymer relaxation time.
Since the latter is conformation-dependent because of hydrodynamic backflow,
it was argued that 
a hysteretic transition between a coiled and a stretched polymer conformation 
may occur upon varying $\kappa$
\cite{Gennes:JCP60-74-5030,Hinch:PF20-77-S22}.
Implicit in the use of the time criterion is the assumption that 
the polymer is in the steady state corresponding to the local flow field 
at any time.
This requires 
variations in the flow field experienced by the polymer to be slow --
an assumption which is probably not valid in flows
as they occur \eg in turbulent drag reduction.
An interesting model for such rapidly varying flows is provided by
the immediate start--up and cessation of simple flows
like the uniform flow acting on a tethered polymer considered here.  
For this case we identify several regimes governed by different decay laws 
which have not been described before.
This complex behavior cannot be captured by a simple dumbbell model
and it coincides with the relaxation of fluctuations in the steady state 
only in the linear response regime of polymer dynamics.

\section{Bead Spring Polymer Model}
In our simulations we follow the Brownian dynamics of
a  bead spring model as described in more detail in Refs.
\cite{Rzehak:NN1-99,Rzehak:NN0-99}.
The  equation of motion for the vector
$\vec{R}= (\vec{R}_1, ..., \vec{R}_N)$  of all bead positions
may be written in the form
\begin{equation}
\label{eq_motion}
\dot{\vec{R}}
=
  \vec{v}
+ \tens{H} \:   \vec{F}
+ \sqrt{2 k_B T \: \tens{H}} \: \mbox{\boldmath $\xi$}
\, .
\end{equation}
Here $\vec{v} = v \, \hat{\vec{x}}$ is the velocity of the unperturbed flow, 
$k_B$ is the Boltzmann constant, $T$ the temperature
and ${\vec{\xi}}(t)$ is an uncorrelated Gaussian white noise with zero mean
and unit variance. 
The potential forces $\vec{F}$ comprise
a repulsive Lennard Jones force
describing the  excluded volume interactions (EVI)
and the next-neighbor bond forces for which we use
either a linear
or the familiar FENE  
({\bf F}inite {\bf E}xtensible {\bf N}onlinear {\bf E}lastic) force law.
The latter provides 
a reasonable approximation of polymers with a fixed bond length
for flow velocities $v \le 0.5$.
Throughout this work we use a chain length of $N=100$ and 
the parameters appearing in the potential forces ${\vec F}$
are the same as in Ref. \cite{Grest:PRA33-86-3628},
where this choice was shown to prohibit bond crossings in the case with EVI.
The hydrodynamic interactions (HI)
are incorporated in the Oseen tensor approximation \cite{Doi:TPD-86},
where the flow perturbations due to all beads simply superimpose
and the additional drag forces can be expressed by
the off--diagonal part of the mobility tensor $\tens{H}$.
We have 
$  \tens{H}_{ij} 
 = \Omega({\vec R}_i - {\vec R}_j)$  
for $i \not = j$
with 
$  \Omega({\vec r}) 
 = ({\vec 1} + {\hat {\vec r}}{\hat {\vec r}}^T) /( 8 \pi\eta |{\vec r}|)$ 
and
$  \tens{H}_{ii} 
 = {\bf 1}/\zeta$ 
with $\zeta = 6 \pi \eta a$
the friction coefficient of 
a bead with radius $a$ in a solvent of viscosity $\eta$.
In contrast to previous work \cite{Larson:PRE55-97-1794,Hatfield:PRL82-99-3548}
we do not average $\tens{H}$
which would effectively linearize the equation of motion.
Since the Oseen tensor becomes non-positive for bead separations of 
the order of $a$, we always consider HI together with EVI.
The values $\eta =0.2$, $\zeta=1.0$, $k_B T =1.0$
and $b=1.0$ (harmonic springs)
or $b=0.961$ (FENE springs) for the bond length $b$
fix the strength of the HI
and the units of force, energy, length etc.~.

\section{Covariance Matrix}
Analytically the polymer relaxation spectrum and modes can be calculated 
only for the Rouse and Zimm models
\cite{Rouse:JCP21-53-1272,Zimm:JCP24-56-269}
which are governed by linear equations of motion
by virtue of the assumptions made.
For realistic polymer models in contrast,
the equation of motion, as described above, is nonlinear.

Nevertheless,
the relaxation times  $\tau_p$ of the Rouse amplitudes
are often used also for nonlinear polymer models in order to
characterize the dynamics of fluctuations in steady states.
Each Rouse amplitude is obtained by 
projecting a time series of polymer conformations 
on the respective Rouse mode 
and $\tau_p$ is determined by 
an exponential fit to the time evolution of
this amplitude \cite{Ahlrichs:JCP111-99-8225}.
The determination of $\tau_p$ in this way is 
tedious and error-prone 
so that in practice only the few longest relaxation times can be obtained. 
In addition,
the significance of the Rouse modes for nonlinear models  
is unclear.
Another approach \cite{Perkins:SCI264-94-822} 
aims at the relaxation spectrum directly,
without referring to corresponding modes,
by applying an inverse Laplace transform 
to a single time series of some observable 
like the end--to--end distance. 
Due to the presence of noise in the data 
this requires the use of special regularization techniques 
and the results may depend strongly on the regularization parameter. 

Here we determine  
the polymer relaxation spectrum 
and a corresponding set of uncorrelated modes 
from the covariance matrix of bead positions. 
This approach is motivated by a relation between 
the initial decay--rate of correlations in
the dynamics of conformational fluctuations in a steady state 
and the second moments of the statistics of bead positions.
This relation may be proved for 
arbitrary polymer models in thermal equilibrium, 
for models without HI, or 
for a preaveraged mobility tensor $\langle \tens{H} \rangle $
by using the the linear response formalism as described e.g. in Ref.\cite{Doi:TPD-86}.
To be specific:
A large sample of data is generated by 
integrating equation (\ref{eq_motion}) numerically over a long time
and storing all bead positions in equal time intervals.
Then from this time series ${\vec x}({\mu}) = \vec{R}( \mu \cdot \Delta t)$
the mean bead positions
$\left< {\vec R} \right> = 
\sum_{\mu =1}^M \, {\vec R}(\mu \cdot \Delta t)/M$
are calculated  
as well as the covariance matrix $\tens{C}$ 
of the deviations $\vec{y} = \vec{R} - \langle \vec{R} \rangle$
\begin{equation}
\tens{C}
=
\langle\, 
   \vec{y}  \vec{y}^T
\rangle
\langle  
  k_{\rm B} T \, \tens{H} 
\rangle^{-1}
\,. 
\end{equation} 
Once $\tens{C}$ is obtained,  
its eigenvalues $\tau_p$ give 
the complete spectrum of relaxation times 
of fluctuations in a steady state.
The eigenvectors $\vec{y}_p^{\alpha}$ of $\tens{C}$,
$\alpha \!=\! x,y,z$, $p \!=\! 1 \ldots N$,  
are uncorrelated 
and form a basis of the  space of polymer conformations 
such that the  deviations $\vec{y}$ may be expressed as 
a linear  combination $\vec{y} = \sum_{p,\alpha} a_p^{\alpha} \vec{y}_p^{\alpha}$
with amplitudes $a_p^{\alpha}$.
At equilibrium $\langle \vec{R} \rangle = 0$ and
the modes  $\vec{y}_p^{\alpha} = \vec{R}_p^{\alpha}$
describe special polymer conformations 
where the position of the $i$--th bead $[\vec{R}_p^{\alpha}]_i$
may be factored in a direction $\hat{\vec{E}}^{\alpha}$ 
and a distance function $R_{pi}$
as $[\vec{R}_p^{\alpha}]_i \!=\! R_{pi} \hat{\vec{E}}^{\alpha}$.
The relaxation spectrum is threefold degenerate due to symmetry. 
In the presence of a uniform flow  
one can distinguish 
one longitudinal mode along the flow direction
and two transverse modes perpendicular to it.
Since $\langle \vec{R} \rangle \neq 0$ away from equilibrium, 
$[\vec{y}_p^{\alpha}]_i$ is not the absolute position of bead $i$ 
when the mode is excited but the {\em deviation} from its mean position. 
The $[\vec{y}_p^{\alpha}]_i$ are a natural generalization of the Rouse modes 
for deformed polymers with nonlinear laws of motion.
From a practical point of view this method 
to compute relaxation times is 
fast,
easy to use,
needs no adjustments of parameters or additional assumptions, 
and yields the complete spectrum in one sweep.
A more detailed discussion and
a possible extension of this method 
that does not invoke preaveraging of $\tens{H}$ 
will be given elsewhere. 

\section{Relaxation of  Fluctuations}
\begin{figure}
\vspace{-10mm}
\hspace{-3mm} 
{\epsfxsize 7.5 cm \epsfbox{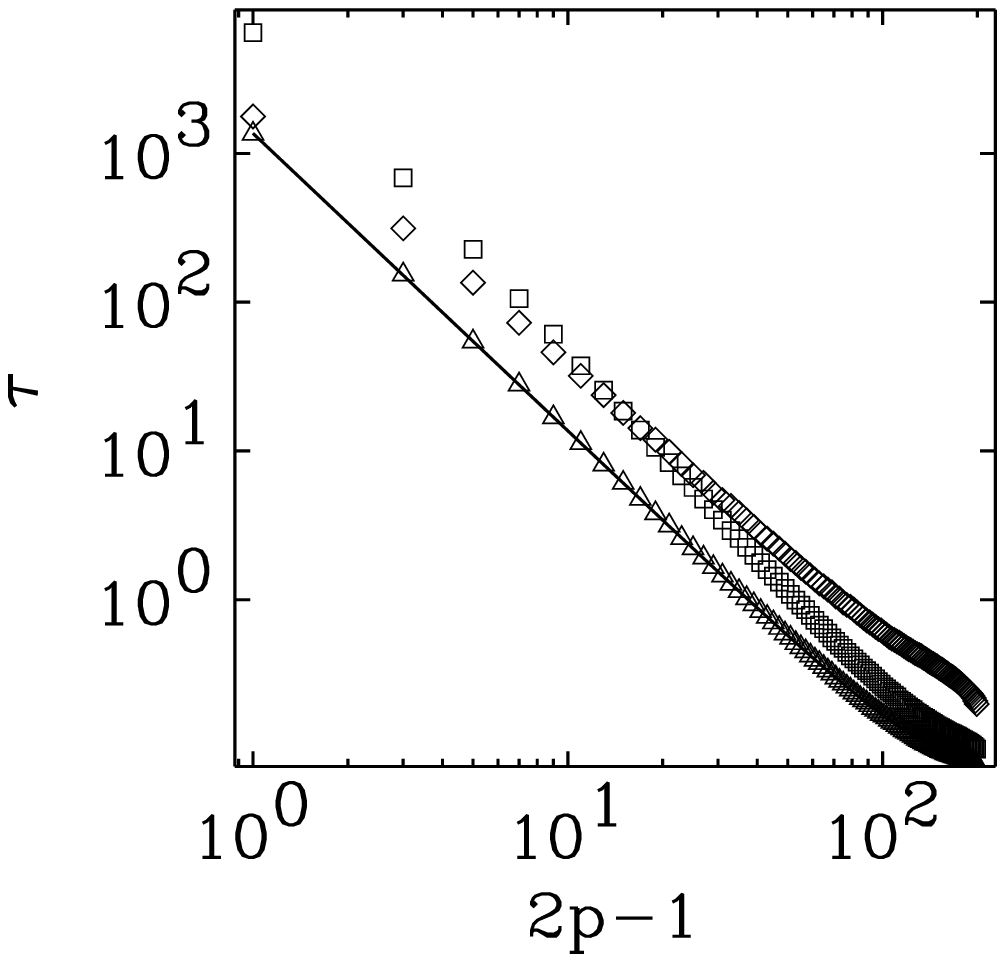}}
\hspace{-8mm} 
{\epsfxsize 7.5 cm \epsfbox{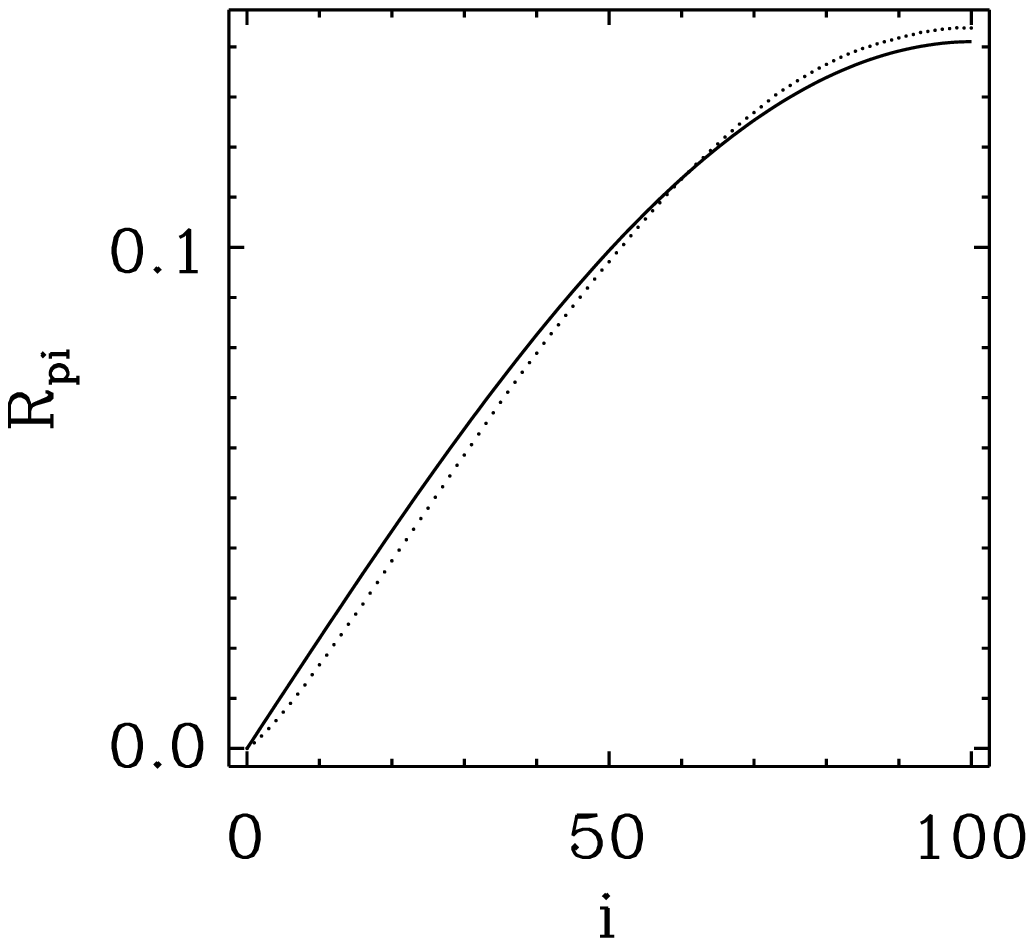}}
\caption{
Left: 
Relaxation spectra for the Rouse chain  
calculated analytically (solid line) and from the matrix $\tens{C}$ (triangles) 
compared with the case including EVI (squares)
or both EVI and HI (diamonds).
Right: First mode ($p=1$) for a pure
Rouse chain (solid line) and with EVI 
included (dotted line).
}%
\label{fig_eq_models_spectrum}
\vspace{-2mm}
\end {figure}

\begin{figure}
\vspace{-10mm}

\hspace{-3mm} 
{\epsfxsize 7.5 cm \epsfbox{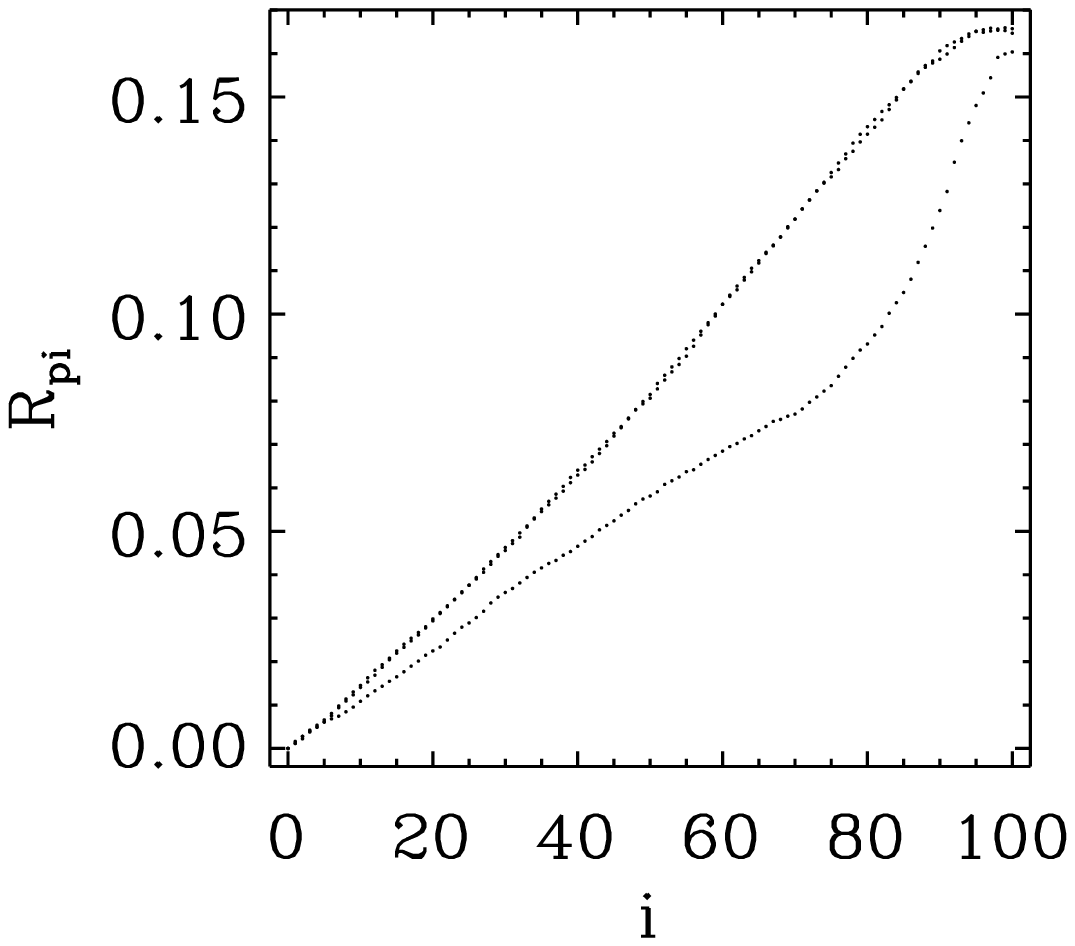}} 
\hspace{-8mm}
{\epsfxsize 7.5 cm \epsfbox{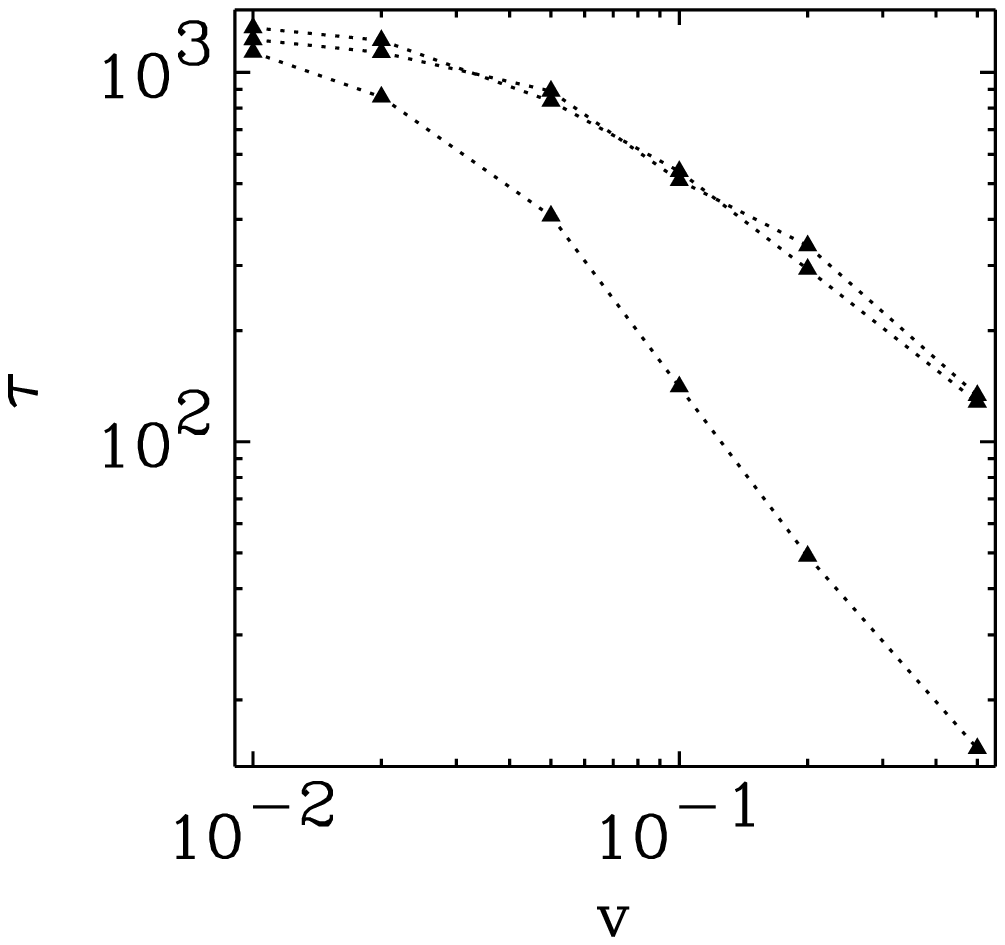}}
\caption{
Relaxation modes and times for a pure FENE chain.
Left: First triplet (p=1) of relaxation modes for $v=0.2$.
Right: Triplet of longest relaxation times as function of the flow velocity $v$
(the dotted lines are only a guide to the eye here).
In each figure, the lower line corresponds to the longitudinal mode
where $\vec{E}^{\alpha}$ is aligned with the flow, 
while the upper two lines correspond to the two degenerate transverse modes.
}
\label{fig_neq_f_modes}
\vspace{-2mm}
\end{figure}

The relaxation spectrum of a Rouse chain in thermal equilibrium 
as calculated by the method 
described above 
is shown in \fig{fig_eq_models_spectrum}.
It agrees perfectly with the 
analytical result
\begin{equation}
\label{rouse_times}
\tau_p^R 
= 
\frac{\zeta}{k_H}\,
\left( 
  4 \sin^2\left( 
            \frac{2p - 1}{2N + 1} \: \frac{\pi}{2} 
          \right) 
\right)^{-1}  
\, ,
\end{equation}
where $\zeta$ is the single bead friction coefficient 
and $k_{\rm H}$ is the force constant of the springs connecting the beads.
The usual scaling
$\tau_p^R \propto (2p-1)^{-2}$ \cite{Gennes:SCPP-81,Doi:TPD-86}
follows for $(2p-1)/N \rightarrow 0$.
Of the nonlinear effects considered here 
EVI have not been included in previous work 
concerned with strong deformations of polymers
\cite{Gennes:JCP60-74-5030,Hinch:PF20-77-S22,Larson:PRE55-97-1794,Hatfield:PRL82-99-3548}.
However, a comparison with the Rouse modes 
in \fig{fig_eq_models_spectrum} shows 
that even in equilibrium EVI leads to 
a small but significant modification of the relaxation modes.
The scaling of the relaxation times in this case is
$\tau_p \propto (2p-1)^{-2.2}$.
This is precisely the inverse of
the power law dependence on the chain length $\tau \propto N^{2\nu+1}$
predicted by scaling theory \cite{Gennes:SCPP-81,Doi:TPD-86}
for a polymer with EVI where $\nu = 3/5$
which is also reproduced by the analysis given above.
In contrast to scaling theory,
our method also yields a number for the longest relaxation time
$\tau_1 = 6.5 \cdot 10^{3}$
which turns out to be about a factor of $5$ larger with than without EVI.
Upon addition of HI to the model the properties 
of the Zimm model \cite{Zimm:JCP24-56-269,Doi:TPD-86}
are reproduced, \ie $\tau_p \propto (2p-1)^{-1.8}$.
%
%
%
%
Replacing harmonic  by FENE springs
does not affect the equilibrium relaxation times or modes.
Away from equilibrium, however, this has a profound impact.
As shown in \fig{fig_neq_f_modes},
the modes are very different from the Rouse form
and the relaxation times decrease strongly with increasing flow velocity.
A decrease of the relaxation times  has been reported recently 
also for a chain which is pulled at {\em the ends} \cite{Hatfield:PRL82-99-3548}.
Experimentally it was observed that the polymer relaxation in this case
can be described by the Rouse modes \cite{Quake:NAT388-97-151}.
Pulling a polymer at the ends, however, is special in that 
the tension along the polymer is everywhere the same.
In contrast,
when the polymer is stretched by a flow
then the tension in general is inhomogeneous \cite{Rzehak:NN1-99}.
A comparison of the modes calculated for both cases reveals that
just this {\it inhomogeneity}
causes the strong deviation of the modes from the Rouse form
as shown in \fig{fig_neq_f_modes}.
We remark that
such inhomogeneous tension also arises due to EVI and HI \cite{Rzehak:NN1-99}.

To identify the effects of HI on the relaxation of a polymer in uniform flow, 
we compare models with harmonic springs and EVI 
for the two cases without and with HI. 
In the first case the relaxation times decrease
in stronger flows since EVI are less important 
for an elongated polymer  \cite{Rzehak:NN1-99}.
For $v \gtrsim 0.2$ the modes assume Rouse shape and both
scaling and numerical values of the relaxation spectrum
exhibit a crossover to Rouse behavior. 
Upon adding  HI to the model, the relaxation times behave non-monotonic:
after an initial decay they increase with $v$ in an intermediate 
range $0.2 \lesssim v \lesssim 2.0$
and only at  very large   flow velocities $v \gt 5.0$
there is a crossover to Rouse behavior.
For the chain lengths available in the present study,
the combination of FENE springs with EVI and HI results in 
a monotonic decrease of the longest relaxation time. 
%
%
%
%
%
%
%
%
%
%
%
%
%

\section{Average Dynamics}
\begin{figure}
\vspace{-10mm}
\hspace{-3mm}
{\epsfxsize 7.5cm \epsfbox{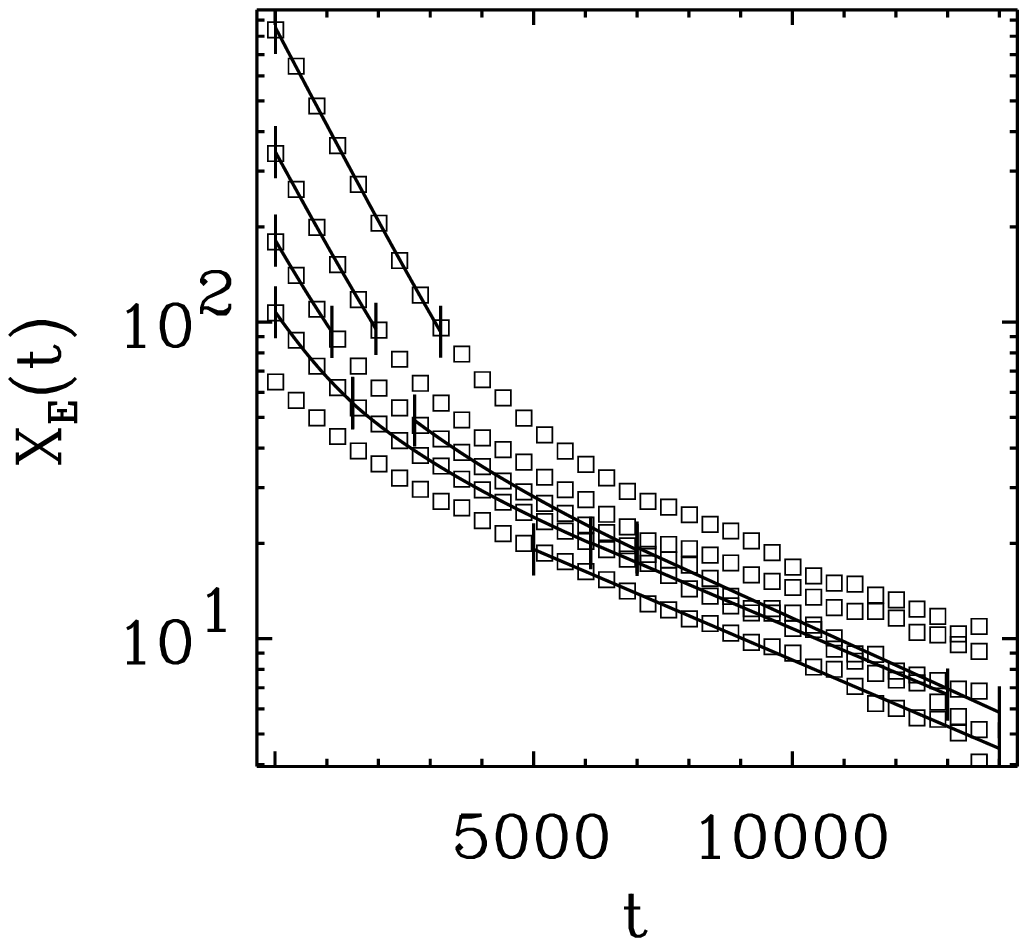}}  
\hspace{-8mm}
{\epsfxsize 7.5cm \epsfbox{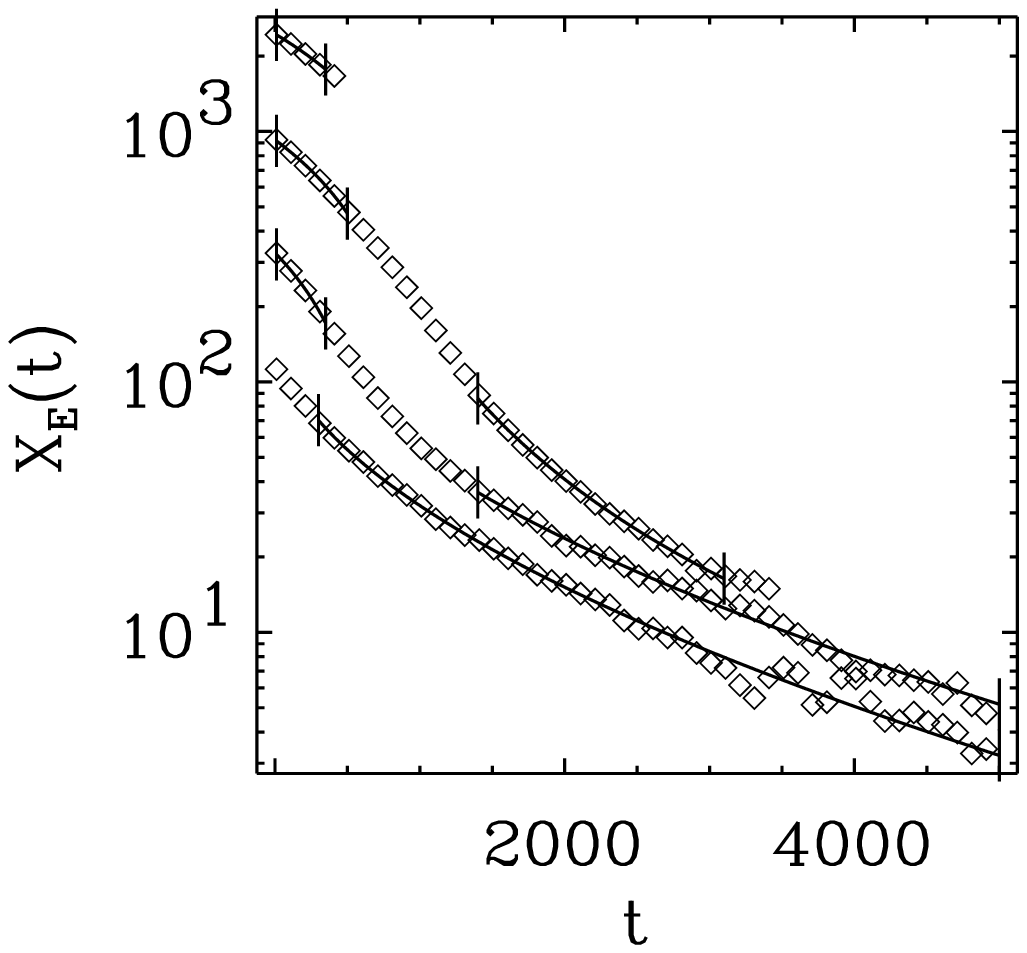}}
\caption{
Evolution of the $x$--component of the end--to--end vector $X_E(t)$
averaged over $100$ runs.
Left: Rouse chain with EVI 
after cessation of a flow with $v=0.5,0.2,0.1,0.05,0.02$ from top to bottom.
Right: Rouse chain with both EVI and HI  
after cessation of a flow with $v=2.0,1.0,0.5,0.2$ from top to bottom.
Different regimes of the relaxation process are marked by vertical bars
on the solid lines representing fits with simple decay laws (see text).
Only every 30-th (squares) respectively 10-th  (diamonds) point of the data 
used for the fits is shown.
}
\label{fig_dyn-_reh}
\vspace{-2mm}
\end{figure}

After cessation of the uniform flow 
the tethered polymer relaxes from an elongated to a coiled state.  
For the model with harmonic springs
the mean $x$--component of the end--to--end vector $X_E(t)$ 
decays exponentially with a time constant $\tau_p^R$, cf. \eq{rouse_times}.
When EVI are added, 
the polymer relaxation exhibits 
three distinct regimes as shown in \fig{fig_dyn-_reh}.
Initially, $X_E(t)$ decays exponentially with time constant $\tau_1^R$. 
This regime is dominated by 
the relaxation of the stretched part of the chain close to the tether-point 
where EVI are negligible. 
Then, one finds a stretched exponential decay 
$X_E(t) \propto \exp(-(t/\tau_1^R)^{\beta})$
with stretching exponent $\beta \simeq 0.5$.
Neither simple exponentials nor power laws 
gave satisfactory agreement with the simulation data.
During this stage, the relaxation is increasingly slowed down because 
self-avoidance constraints become more important. 
A common explanation of stretched exponential behavior is by
a superposition of single exponential decays
with time constants that obey a power law.
A power law spectrum of modes is provided here by the Rouse modes
which {\em all} become excited due to the nonlinear coupling by EVI
even when initially only the lowest mode is present.
It is unlikely that this complex relaxation behavior can be captured by 
a single mode theory like the dumbbell model.
In a third regime finally, which sets in when 
$X_E(t)$ equals the equilibrium end--to--end distance $R_E = 22.2$,
the entangling process is completed 
and the vanishing equilibrium value of $X_E$ is approached by
only local rearrangements within the coil.
This process is again described by a single exponential decay law
where the time constant is now given by 
the longest  relaxation time  in equilibrium, cf. \fig{fig_eq_models_spectrum}.
When in addition to EVI also HI are included in the model,
a very different behavior results 
in the initial phase of the relaxation
for large velocities $v \ge 0.5$.
In this case the polymer at first appears to move freely, 
\ie $\partial X_E / \partial t \propto v$.
This is the consequence of 
a large flow perturbation
due to {\em average} forces acting on the beads
which leads to a highly cooperative motion.
This is very different from the steady state
where the flow perturbation is caused only by {\em fluctuating} forces.
Replacing the harmonic by FENE springs
causes a short and rapid initial recoil
but the later stages of the relaxation remain qualitatively unaffected.

After the flow is started, simulation data (not shown in the figure)
reveal for all models at first a short period of apparent free polymer motion.
This is followed by 
a single exponential growth of $X_E$
which is well described by 
the longest relaxation time in the stretched final state.
The pure Rouse model is exceptional since
the initial regime is confined to the shortest relaxation time.
%
%
%
%

\section{Conclusions}
Using the second moments of the statistics of bead positions, 
we have applied a powerful method for 
the determination of the complete polymer relaxation spectrum 
and corresponding modes
from simulation data. 
Our approach works for 
both nonlinear polymer models
and arbitrary steady states 
as caused by the action of flows or forces on the polymer. 
For a tethered polymer in uniform flow,
the finite extensibility causes
a monotonic {\em decrease} of relaxation times with increasing elongation.
Older theories of the coil stretch transition in elongational flow,
in contrast, were based on the assumption of 
{\em increasing} relaxation times \cite{Gennes:SCPP-81}.
This finding agrees with 
recent results obtained for a chain pulled only at the ends 
\cite{Hatfield:PRL82-99-3548,Quake:NAT388-97-151}
where the modes remain Rouse-like.
The modes for a polymer stretched by a flow, in contrast, 
are very different from the Rouse modes
because the tension along the polymer then in general is inhomogeneous. 
On the other hand we found that HI may lead to 
a non-monotonic velocity dependence of the relaxation times.
In order to draw a final conclusion about 
the occurrence of a coil-stretch transition as predicted for elongational flow,
longer chains would have to be considered. 

Going beyond the relaxation dynamics of polymers close to steady states,
we also dropped the common preaveraging approximation for the mobility tensor. 
Thus, in contrast to other recent work, all
effects relevant for uncharged polymers in flows are covered. 
Specifically, we  have considered the response of a tethered polymer 
to start-up or cessation of the flow and found that 
EVI causes a stretched exponential relaxation 
which can only be described by 
a model with many degrees of freedom 
and not by a simple dumbbell model.
Scaling relations predicted by common blob models
\cite{Brochard:EL26-94-511+Brochard:EL30-95-387+Marciano:MM28-95-985} 
are unlikely to occur in simulations
because the blob models do not include inhomogeneous effects of EVI and HI
\cite{Rzehak:NN2-99,Rzehak:NN1-99}.
A more detailed discussion of relations to other work 
and an extension of this study to other flow fields will be
given elsewhere. 

%
%
\stars

It is a pleasure to acknowledge fruitful discussions with 
B. D\"unweg and D. Kienle.

%
%
%

\end{document}